\documentclass[12pt]{article}
\usepackage{amsmath,amsthm,amscd,amsfonts,amssymb}
\usepackage{latexsym}
\newcommand{\CC}{\mathbb C}
\newcommand{\DD}{\mathbb D}
\newcommand{\PP}{\mathbb P}
\newcommand{\RR}{\mathbb R}
\newcommand{\ZZ}{\mathbb Z}
\newcommand{\E}{\mathbb E}
\newcommand{\Znu}{\ZZ^\nu}
\newtheorem{thm}{Theorem}[section]
\newtheorem{lemma}[thm]{Lemma}

\newtheorem{prop}[thm]{Proposition}
\newtheorem{defin}[thm]{Definition}

\newtheorem{assum}[thm]{Assumptions}
\begin{document}
\title{ Localization and mobility edge for sparsely random potentials }
\author{M Krishna \\ 
Institute of Mathematical Sciences\\ Taramani,
Chennai 600 113, India \\
J Obermeit\\
Wittener Str, 220\\
44803, Bochum, Germany}
\date{4 November 1998}
\maketitle
\begin{abstract}
In this paper we consider sparsely random potentials 
$\lambda V^\omega$, $V^\omega$ supported on a sparse subset S 
of $\ZZ^\nu$ and a
bounded self-adjoint free part $H_0$ and show the presence of absolutely
continuous spectrum and pure point spectrum for $H_0 +
\lambda V^\omega$ when $\lambda$ is large and $V^\omega(n)$ are
independent random variables for n in S, with either identical
distributions or distributions whose variance goes to $\infty$ with n,
when $\nu \geq 5$.
\end{abstract}

\section{Introduction}
In this paper we consider random potentials on the $\nu$
dimensional lattice.  The class of potentials considered in this paper
are with sparse support but large disorder. A set S is sparse for us
if the number of sites in any cube that belong to S grows at a
fractional power of the volume of the cube, as the volume of the cube
goes to $\infty$.  The criterion of
sparseness is stated in terms of the decay of the wave packets
generated by the free evolution in our assumption on sparseness. 

We are motivated by the difficult problem of existence of absolutely
continuous spectrum in the higher dimensional Anderson model with
small disorder and especially the question of whether a sharp mobility edge
exists or not.  We look for some models of random potentials
that exhibit this behaviour.

In an earlier version of this paper we claimed that there is a sharp
mobility edge for the large disorder models studied in this paper, 
but the argument we presented there had a gap.

However the techniques of Aizenman-Molchanov allow for the inclusion
of a potential going to infinity at $\infty$, for which the claim on
the mobility edge is true. This is our second theorem. 

We list a part of the relevant literature on the Anderson model
in the references for the benefit of the reader.

Once class
of models considered in Krishna \cite{mk1} were random potentials with decaying
randomness with independent potentials at different sites, 
in higher dimensions,  for which  some absolutely
continuous spectrum was shown to exist.  
For these class of models the question of mobility edge was considered
in Kirsch-Krishna-Obermeit \cite{kko}, and the answer was obtained for
a class of potentials, when the decay rate is fast and the dimension
is large. 

In this paper we use the criterion from scattering theory for showing the existence
of absolutely continuous spectrum and the 
technique of Aizenman-Molchanov \cite{am} to verify the Simon-Wolff
\cite{sw} criterion for the absence of continuous spectrum 
outside a band.

Firstly we assume that the unperturbed part satisfies the
\begin{assum}
\label{ass1}
Let $H_0$ be a bounded self-adjoint operator on $\ell^2(\ZZ^\nu)$,
with 
\begin{equation}
\begin{split}
\|H_0\|_{s} \equiv \sup_{m} ~~ 
\left(\sum_{n \in \ZZ^\nu} |\langle \delta_n, H_0
\delta_m\rangle|^{s}\right)^{1/s} ~~ < \infty\\
\end{split}
\end{equation}
for all $s_0< s < 1$ with $s_0 >0$.  
\end{assum}

We note that in the case when $H_0 =\Delta$, the usual finite
difference operator given by $(\Delta u)(n) = \sum_{|n -i| =1}
u(i)$, we have $\|H_0\|_s = (2\nu)^{1/s}$.
We also remark that once we
assume the finiteness of the sum for $s_0$ it follows for all s,
however we used the definition to fix the notation $\|H_0\|_s$.

Next we consider some subsets of $\ZZ^\nu$ which are regular in the
following sense with respect to $H_0$.  This criterion is useful in
proving the existence of the wave operators later.

\begin{defin}
\label{defi1}
Let S be any subset of $\Znu$ and $P_S$ the orthogonal projection on
to the subspace $\ell^2(S)$ in $\ell^2(\ZZ^\nu)$. 
 Then we call S {\bf sparse relative to $H_0$},
 a self-adjoint operator with non-empty absolutely
continuous spectrum, whenever there exists a dense subset 
$\DD$  contained in the absolutely continuous subspace of $H_0$
such that
\begin{equation}
 \int dt \|P_S exp\{-itH_0\}\phi\| 
< \infty, ~~ \forall \phi \in \DD.
\end{equation}
If A is an operator of multiplication by a real sequence $\left\{a_n, n \in
\ZZ^\nu\right\}$, then we say
S is sparse relative to $H_0$ with weight A if,
\begin{equation}
 \int dt \|A P_S exp\{-itH_0\}\phi\| 
< \infty, ~~ \forall \phi \in \DD.
\end{equation}

\end{defin}
{\bf Remark:}  1. The assumption may not be satisfied
even for finite sets S if $H_0$ is an arbitrary
self-adjoint operator with non-empty absolutely continuous spectrum.
For certain class of S with infinite cardinality and for $H_0 = \Delta$,
and  dimension $\nu \geq 4$, the $H_0$ sparseness  
was shown in Krishna \cite{mk2}.  Such sets S would be sparse in $\Znu$.
The class of subsets considered there are bigger than those
considered as examples below.

2. One should contrast the sparseness criterion with the similar
looking smoothness criterion widely used in scattering theory.
 
3. We note that there cannot be any non-zero operator $H_0$ on
$\ell^2(\ZZ^\nu)$ with some
absolutely continuous spectrum, such that $k\ZZ^\nu$ is sparse
relative to it for any non-zero integer k.

Finally we assume that
the random potential has sparse support and that its distribution has
finite variance.  The
finiteness of the variance is needed in our proof of the presence of
absolutely continuous spectrum for the perturbed operators.

\begin{assum}
\label{ass2}
Let S be any subset of $\ZZ^\nu$. Let $V_S^\omega(n), n \in S$
be independent real valued random variables which are
identically distributed
according to an absolutely continuous probability  
distribution $\mu$ on $\RR$ satisfying 
\begin{equation}
\begin{split}
\sigma^2 \equiv \int |x|^2 d\mu(x) < \infty ~~ \text{and} ~~
 \mu(a-\delta, a+\delta) \leq C
\delta \mu(a - b, a + b), ~~ \forall  a\\
  ~~ \text{and} ~~ 0 \leq \delta
< 1,  
\end{split}
\end{equation}
with C independent of a and $\delta$ for some $b \geq 1$.    
\end{assum}

(We denote by $\mu_0$ the atomic probability measure on $\RR$ 
giving mass 1 to the point 0.  Then
 $V_S^\omega(n)$ are real
valued measurable functions on a probability space  $(\Omega, \PP)$,
$\Omega = \RR^S \times \RR^{S^c}$ and $\PP = \times_S (\mu) \times 
\times_{S^c}(\mu_0)$, so the parameter $\omega$
denotes a point in $\Omega$ -- or equivalently a real valued sequence
indexed by points of $\ZZ^\nu$ -- 
and all our future references to a.e.,  are denoted with respect to
this measure $\PP$.)

Let $V_S^\omega$ denote the operator 
$$
(V_S^\omega u)(n) = V_S^\omega (n) \chi_S(n)u(n), ~~ n \in \ZZ^\nu.
$$
where $\chi_S$ denotes the indicator function of the set S. 
Let $H_0$ be some non random background bounded self-adjoint operator as in
assumption (\ref{ass1}).  In the case when the measure $\mu$ has
infinite support, the above are a family of unbounded self-adjoint
operators having the set of finite vectors in their domain for almost
every $\omega$.  

The main theorems  of this paper are the following.
\begin{thm}
\label{thm1}
Let
$H_0$ be any bounded self-adjoint operator satisfying assumption
(\ref{ass1}) and having some absolutely continuous
spectrum.  Let $S \subset \Znu$ be  sparse relative to $H_0$. 
Let $V_S^\omega(n) $  satisfy the assumption
(\ref{ass2}) and if $\mu$ has infinite support assume further that 
\begin{equation}
\label{eqthm}
\sum_{m \in S} |(1 + |m|)^\beta (e^{-it H_0}\phi)(m)|^2 < \infty, \phi \in \DD, 
\end{equation}
for some $\beta > \nu$ and each fixed t.  Consider the operator
$H^\omega_\lambda = H_0 + \lambda V_S^\omega$.
 Then, 
\begin{enumerate}
\item $\sigma_{ac}(H^\omega_\lambda) \supset \sigma_{ac}(H_0) $ a.e.
$\omega$  and

\item For each $0 < s < 1$, there is a $\lambda_s < \infty$,
such that for any $\lambda > \lambda_s$, 
$$
 \sigma_c(H^\omega_\lambda) \subset
 [-\|H_0\|_s, \|H_0\|_s].
$$
\end{enumerate}
\end{thm}

{\bf Remark:}
1. The above theorem does not show the existence of spectrum outside
$[-\|H_0\|_s, \|H_0\|_s]$.  The existence of spectrum there can be
shown for large $\lambda$ on the lines of Kirsch-Krishna-Obermeit \cite{kko},
even for the case when $\mu$ has compact support.  The technique uses
rank one perturbations to obtain a Weyl sequence.

2. The proof of the above theorem relies in part on the decoupling
bounds obtained by Aizenman-Molchanov \cite{am}, where the  
constants $\lambda_s \rightarrow \infty$ as s approaches 1, so for
any finite coupling constant $\lambda$, there is always an 
$s < 1 $, with $\lambda < \lambda_s$. Therefore for any finite
$\lambda$ there is always a region  $(\|H_0\|_1, \|H_0\|_s) \cup
(-\|H_0\|_s, -\|H_0\|_1)$, where we cannot determine the spectral
behaviour by this method.  However when the potential goes to $\infty$
at $\infty$, while being supported on the sparse set, this problem can
be avoided, which is our next theorem.

3. Recently considering surface randomness, Jaksic -Molchanov \cite{jm}
proved pure point spectrum outside [-4,4] in the case when the randomness is on the
boundary of  $\ZZ^2_+$ and this is not a sparse set.

\begin{thm}
\label{thm2}
Let
$H_0$ and $V_S^\omega$ be as in theorem (\ref{thm1}).  Let $a_n$ be a
sequence of positive numbers $|a_n| \rightarrow \infty$ as $|n|
\rightarrow \infty$.  Let A be the operator of multiplication by $a_n$
and let $V_{S,A}^\omega = A V_S^\omega$. Assume further that S is
sparse relative to $H_0$ with weight A.  Consider $H^\omega = H_0 + AV_S^\omega$.  
 Then, 
\begin{enumerate}
\item $\sigma_{ac}(H^\omega) \supset \sigma_{ac}(H_0) $ a.e.
$\omega$  and

\item The continuous spectrum satisfies 
$$
 \sigma_c(H^\omega) \subset
 [-\|H_0\|_1, \|H_0\|_1].
$$
\end{enumerate}
\end{thm}

{\bf Remark:} 1. In the case of operators $H_0$ the boundary of whose
spectrum coincides with the points $\pm \|H_0\|_1$, they are the
mobility edges .  This happens for example for the operators $\Delta$
or some of the other examples given later.

{\bf Acknowledgment:} We thank Prof Werner Kirsch for helpful
discussions and encouragement.  We also thank an anonymous referee for
pointing out (indirectly) the errors in an earlier version.  JO wishes
to thank the Institute of Mathematical Sciences for a visit when most
of this work was done.

\section{Proof of the theorems}

Before we get to the proof of the theorems, we explain the motivation
for considering the above model and the ideas
involved briefly for the benefit of the reader.  
Our motivation comes from the effort to find random models that
exhibit the expected behaviour of the spectrum of the Anderson model
at small disorder.  The reader who does not wish to be lost in the
generality, could consider $H_0 = \Delta$ and work through the paper.

The general theory
of scattering gives a way of checking if a given self-adjoint operator
A has some absolutely continuous spectrum or not.  The technique is to
find another self-adjoint operator B which is known to have some
absolutely continuous spectrum and verify that the wave operators
$ S-\lim e^{-itA} e^{itB}P_{ac}(B)$ exist.  Then it follows that
$\sigma_{ac}(A) \supset
\sigma_{ac}(B)$.  We use this technique with B as the free operator $H_0$
of our model and A the random operator.  To show the existence of the above
limits is done via the Cook method of showing that the integral
$\int \|(A - B)e^{itB}f\| dt$  is finite.  The sparseness condition we
imposed is an abstract condition that requires that the support (in 
$\ZZ^\nu$) of the potential is  such as to make this integral finite,
for a set of f dense in  the absolutely continuous spectral subspace of
B.

Our class of examples come from operators of multiplication by real
valued functions on $[0, 2\pi]^\nu$, so that they
commute with $\Delta$ and also satisfy
some smoothness condition so as to make the matrix elements $H_0(n,m)$
in $\ell^2(\ZZ^\nu)$ decay at a required rate, in addition to giving
decay in t for the function $e^{itB}(n,m)$, obtained via a stationary
phase argument. 

We impose the condition
on the support of the
random potential based on the behaviour of the integrals 
$(e^{itB}f)(m)$ as a function of m and t, so as to make them sparse
relative to B {\it \'{a} la} the sparseness assumption.

As for the pure point spectrum outside the smallest interval
containing the absolutely continuous spectrum of our models, we use
the Aizenman-Molchanov technique.  The idea behind the method is to
use the presence of randomness with large coupling to obtain decay
estimates on the averages of small moments of the resolvent kernels
of the random operators.  Aizenman-Molchanov estimates use the
fact that the potentials have components which are mutually
independent at different sites 
and also the regularity of their distribution.
However this technique fails at the points
in $\ZZ^\nu$, where the random potential is absent.  So we 
modify the method, and use their estimates at sites in the
lattice where the potential is non-zero and at the sites where the
potential is zero, we use the fact that $|E|$ is large.  Thus we use
both the large $\lambda$ and large $|E|$ conditions in obtaining the
exponential decay estimates on the Green functions.

{\bf Proof of theorem (\ref{thm1}):} As for item (1) of the theorem 
we prove that the wave operators
namely 
\begin{equation}
\text{S-lim} ~~ exp\{iH^\omega_\lambda t\}exp\{-iH_0 t\}P_{ac}(H_0)
\end{equation}
exist, 
where $P_{ac}(H_0)$ denotes the orthogonal projection onto the
absolutely continuous spectral subspace of $H_0$.  That this
implies (1) is standard, see \cite{mk1}).  

Suppose we have, 
$\int \|V_S^\omega e^{-itH_0}\phi\| dt < \infty$ for
almost every $\omega$ for all $\phi \in \DD$, then the sequence
$e^{itH^\omega_\lambda}e^{-itH_0}\phi$ is Cauchy for all 
$\phi \in \DD$, from which the existence of the wave operators follows,
since $\DD$ is dense in the absolutely continuous spectral
subspace of $H_0$, by assumption.

We take $\beta$ as in the theorem and consider the sets 
$$
A_m = \{\omega : |V^\omega(m)| > (1 + |m|)^\beta\}, ~~ m \in S.
$$
Then by assumption (\ref{ass2}), on the variance of $\mu$, we have 
$$
\sum_{m \in S} \PP(A_m) \leq 
\sum_{m \in S} \int_{A_m} \frac{1}{x} x d\mu(x) \leq \sigma
\sum_{m \in \ZZ^\nu} 1/(1+|m|)^\beta < \infty,
$$
using the Cauchy-Schwarz inequality.
Therefore by Borel-Cantelli lemma, $\Omega_0 = \{\omega : \sum_{m \in
S} |m|^{2\beta} |u(m)|^2 < \infty, ~~ \text{for all} ~~ u \in Dom(V_S^\omega)\}$ has
probability 1. 

Then it follows from the assumption in the theorem,  equation (\ref{eqthm}),
that the random variable $\|V_S^\omega e^{-itH_0}\phi\|$, is defined finitely
on $\Omega_0$ for each t fixed. 

The square of this random variable is integrable in $\omega$ for each fixed t, as
can be seen from the following estimate, $H_0$ sparseness of S and use 
of Fubini's theorem.
\begin{equation}
\begin{split}
\E \{ \|V_S^\omega e^{-itH_0}\phi)\|^2\} \\
\sum_{m \in S} \E \{ |V^\omega(m)|^2 |e^{-itH_0}\phi)(m)|^2\} \\
\leq \sum_{m \in S} \E \{ |V^\omega(m)|^2\} |e^{-itH_0}\phi)(m)|^2
\leq {\sigma}^2\sum_{m \in S} |e^{-itH_0}\phi)(m)|^2.
\end{split}
\end{equation}
Therefore its average value is also integrable in t as can be seen from the
inequalities,
\begin{equation}
\begin{split}
 \int dt \E \{ \|V_S^\omega e^{-itH_0}\phi\|\} \leq 
 \int dt \{ \E  \|V_S^\omega e^{-itH_0}\phi\|^2 \}^{1/2}\\
 \leq  \sigma \int dt  \|P_S e^{-itH_0}\phi\|   <\infty
\end{split}
\end{equation}
with the last inequality resulting by assumption of the $H_0$ sparseness of
S.  Then using Fubini, we conclude that 
$$
  \E \{ \int dt \|V_S^\omega e^{-itH_0}\phi\| \} < \infty
$$
showing that $\|V_S^\omega e^{-itH_0}\phi\|$ is integrable in t for 
almost all $\omega$, proving the result.

The proof of (2) of the theorem follows from the lemma (\ref{lem2}) below.

In the following we
denote by $G(E+i\epsilon, n, m) = \langle \delta_n, (H^\omega_\lambda - E -
i\epsilon)^{-1}\delta_m\rangle$.

\begin{lemma}
\label{lem2}
Suppose $H^\omega_\lambda$ be an operator as in theorem
(\ref{thm1}).  Then  for any $s_0 < s < 1$, there
is a $\lambda_s > 0$ such that for any $\lambda > \lambda_s$ and
a.e. $\omega$ we have $\sigma_c(H^\omega_\lambda) 
\subset [- \|H_0\|_s, \|H_0\|_s]$.
\end{lemma}

{\bf Proof:} 
We prove this lemma by proving that there is a $\lambda_s$ such
that for each $s_0 < s < 1$, the estimate 
\begin{equation}
\label{ieq}
\sum_{m \in \ZZ^\nu} \E\{ |G(E+i\epsilon,n,m)|^s \}
 \leq C < \infty
\end{equation}
is valid whenever $ |E| > \|H_0\|_s$ and $\lambda >
\lambda_s$ with C independent of $\epsilon$.
This estimate implies by integrating over E in an
interval
$[a,b] \subset  (-\infty, -\|H_0\|_s) \cup (\|H_0\|_s , \infty)$ 
and using the fact that $(\sum x_i )^s \leq \sum
x_i^s$ for $x_i \geq 0$ and $0 < s < 1$,
\begin{equation}
\begin{split}
\int_a^b dE ~~  \E \{\sum_{m \in \ZZ^\nu}  |G(E+i\epsilon,n,m)|^2\}^{s/2} \leq \\
\int_a^b dE ~~  \E \{\sum_{m \in \ZZ^\nu}  |G(E+i\epsilon,n,m)|^s\} < \infty.
\end{split}
\end{equation}
Hence for a.e. $(\omega, E) \in \Omega \times [a,b]$, we have
$$
\sum_{m \in \ZZ^\nu} |G(E+i0, n, m)|^s < \infty ~~\text{and hence}
~~
\sum_{m \in \ZZ^\nu} |G(E+i0, n, m)|^2 < \infty.
$$ 
by means of Fatou's lemma and the existence of the limit
$\lim{\epsilon \downarrow 0} 
\sum_{m \in \ZZ^\nu} |G(E+i\epsilon, n, m)|^2 .$
Therefore by the Simon-Wolff \cite{sw} criterion,  the spectral
measure of the operator $H_\lambda^\omega$ associated with
 the vector $\delta_n$ has no
continuous component  supported in $[a, b]$.  Since this happens for
all $n \in \ZZ^\nu$ and since the collection $\{\delta_n\}$ forms an
orthonormal basis in $\ell^2(\ZZ^\nu)$ as n varies in $\ZZ^\nu$, it
follows that
$\sigma_c (H^\omega_\lambda) \cap [a, b] = \emptyset$, for almost all points in
$\Omega$. By taking a countably many bounded intervals we see that
this implies that for almost all points in $\Omega$, 
$\sigma_c (H^\omega_\lambda) \subset [\|H_0\|_s, \|H_0\|_s]$.

Therefore we prove the estimate in equation (\ref{ieq}), to do which
we fix some s in $(s_0,1)$ and consider the equation
\begin{equation}
\begin{split}
(\lambda V_S^\omega(m) - E -i\epsilon)G(E+i\epsilon, n, m) +
\\
\sum_{k\in\ZZ^\nu} \langle\delta_k, H_0 \delta_m\rangle
G(E+i\epsilon, n, k) = \delta_{n.m}.
\end{split}
\end{equation}
We transfer the sum involving $H_0$ to the right hand side and 
take the average of the absolute value raised to power s to
get the inequality, (using the fact that $(\sum x_i )^s \leq \sum
x_i^s$ for $x_i \geq 0$ and $0 < s < 1$),
\begin{equation}
\begin{split}
\E \{ |(\lambda V_S^\omega(m) - E -i\epsilon)
G(E+i\epsilon, n, m)|^s\} \leq \\ \delta_{n,m} + 
\sum_{k\in\ZZ^\nu} |\langle\delta_k, H_0 \delta_m\rangle|^s
\E \{|G(E+i\epsilon, n, k)|^s\}.
\end{split}
\end{equation}
In the inequality below 
we set
$$
C(E, \lambda, s) = 
\begin{cases} |E|^s, & ~~ m \notin S ~~ \text{and}\\
  C(\lambda, s) & 
~~ m \in S
\end{cases}
$$
where $C(\lambda, s) = |\lambda|^s (1 -s )^s D(s)$
is the constant appearing in
proposition (\ref{aprop1}).  Therefore when $|E| > \|H_0\|_s$
and $\lambda > \lambda_s (= \|H_0\|_s/(1 - s)D(s)^{1/s})$,
we can make $C(E, \lambda, s) > \|H_0\|_s^s$.

Now we use the decoupling principle (proposition (\ref{aprop1}))
and Fubini to  interchange the sum and the integral to get,

\begin{equation}
\begin{split}
C(E, \lambda, s)\E \{ G(E+i\epsilon, n, m)|^s\} \leq \\ 
\E \{ |(\lambda V_S^\omega(m) - E -i\epsilon)
G(E+i\epsilon, n, m)|^s\} \leq \\
\delta_{n,m} + 
\sum_{k\in\ZZ^\nu} |\langle\delta_k, H_0 \delta_m\rangle|^s
\E \{|G(E+i\epsilon, n, k)|^s\}.
\end{split}
\end{equation}

Then by our choice of the $\lambda$, $C(E, \lambda, s) >
\|H_0\|_s^s$, so using the proposition below on the  bounds on $\E \{
|G(E+i\epsilon, n,m)|^s\}$ , uniform in $\epsilon$, the proof now follows
on the same lines of Aizenman-Molchanov \cite{am}, by repeating the
above estimate $|n-m|$ times to get the following bound, where we set
\begin{equation}
\label{defks}
\begin{split}
K_s^{*j}(n) = \sum_{i_1, \cdots, i_{j-1},m \in \ZZ^\nu}
\frac{|\langle \delta_{n}, ~ H_0 \delta_{i_1} \rangle |^s 
|\langle \delta_{i_1} , ~H_0 \delta_{i_2} \rangle |^s 
\cdots 
|\langle \delta_{i_{j-1}} , ~H_0 \delta_{m} \rangle |^s}
{C(E,\lambda, s)^j} \\
\text{and} ~~ 
k_s = \|H_0\|_s^s/C(E,\lambda,s) ,
\end{split}
\end{equation}
then $\sup_n K_s^{*j}(n) \leq k_s^j$.
\begin{equation}
\begin{split}
\sum_{m \in \ZZ^\nu} \E \{ G(E+i\epsilon, n, m)|^s\} \leq
1 + \sum_{j = 1}^\infty K_s^{*j}(n)  + \sum_{m \in \ZZ^\nu} K_s^{|n-m|}
 D(E, \lambda, s)\\  
\leq \sum_{j = 0}^\infty k_s^j  + (\sum_{l = 1}^\infty l^{\nu -1}k_s^l)
 D(E, \lambda, s) < \infty,  
\end{split}
\end{equation}
since $k_s <1$ by assumption on E and $\lambda$, with the bound 
independent of $\epsilon$.
(We note that one could have used a Combes-Thomas type argument to
avoid using the uniform bounds provided, the quantities $|\langle
\delta_n H_0\delta_m \rangle|$ have exponential decay in $|n-m|$ as it happens for 
$\Delta$ and other examples with finite range off diagonal parts.)

The uniform bounds below are    analogous to the uniform
bounds obtained by Aizenman-Molchanov \cite{am}(equation (2.12)). 
(The following proposition uses ideas similar to   
the Wegner estimate of Kirsch \cite{kw} or Obermeit
\cite{jo} in the proof of localization.)
 
\begin{prop}
\label{prop1}
Consider the operator $H^\omega_\lambda$ as in theorem
(\ref{thm1}) or $H^\omega$ as in (\ref{thm2}) and let $s_0$ as in
assumption (\ref{ass1}). 
\begin{enumerate}
\item In the case of theorem (\ref{thm1}),
for all 
$E \in \RR \setminus \sigma(H_0)$ and $s_0 < s < 1$,
\begin{equation}
\E\{ |G(E+i\epsilon, n, m)|^s\} \leq D(E, \lambda, s) < \infty.
\label{est1}
\end{equation}
\item In the case of theorem (\ref{thm2}),
for all 
$E \in \RR \setminus \sigma(H_0)$ and $s_0 < s < 1$,
\begin{equation}
\E\{ |G(E+i\epsilon, n, m)|^s\} \leq D(E, \sqrt{|a_m a_n|}, s) < \infty.
\label{est2}
\end{equation}
\end{enumerate}
The constants $D(E, \cdot, s)$ appearing above are uniformly bounded in
E, for E in any compact subset of $\RR \setminus [-\|H_0\|_s,
\|H_0\|_s]$.
\end{prop}

{\noindent \bf Proof:}
We split the proof of part (1) of the proposition in to three cases. 
The proof of part (2) of the theorem is similar, by replacing 
-- for each $n, m \in S$ -- $\lambda$ by $\sqrt{|a_m a_n|}$ (which is
the form in which the estimate of equation (\ref{estam}) is valid, see
for example \cite{kko}, where the estimate was shown to be valid
with $\lambda$ replaced by $a_n$ and $a_m$ seperately, so
by interpolation the present estimate comes out), 
in the estimate of Case 1,
below and going through the proof of all the cases.

{\noindent \it Case 1}: n, m $\in S$.  Then the bound
\begin{equation}
\label{estam}
\E\{ |G(E+i\epsilon, n, m)|^s\} \leq (2\sqrt{2})^s/(|\lambda|^s(1-s)) 
\end{equation}
is similar to the estimate 
proved using  Theorem II.1 
Aizenman-Molchanov \cite{am}.  We designate the constant on the right hand
side of the above inequality as $D_0(E, \lambda, s)$.

{\noindent \it Case 2}: $n \in S$ and $m \in S^c$ or $n \in S^c$ and
$m \in S$.  We consider the possibility $n \in S^c$ and $m \in S$, the
proof of the other possibility is similar.
Let $H_{0, S^c} = P_{S^c}H_0P_{S^c} $,  then $\|H_{0,S^c}\| \leq
\|H_0\|_s$.
We then consider the operator
$H_\lambda^\omega(S) = H_{0,S^c} + V_S^\omega$, set $z = E +
i\epsilon$ and use the resolvent equation
to write
\begin{equation}
\begin{split}
(H_\lambda^\omega - z)^{-1}(n,m) = 
(H_\lambda^\omega(S) - z)^{-1}(n,m) \\
-    \sum_{k \in S^c, l \in S} (H_\lambda^\omega(S) - z)^{-1}(n,k)\\
[(H_0 -
H_{0,S^c})(k, l)] 
(H_\lambda^\omega(S) - z)^{-1}(l,m)  
\end{split}
\end{equation}
where we have used the fact that since $n \in S^c$, when $k \in S$,
 $(H_{0,S^C} -z)^{-1}(n,k) = 0$ and hence $(H_\lambda^\omega(S) - z)^{-1}(n,k) =
0$.  Then it follows that for any $s_0 < s < 1$, the estimate
\begin{equation}
\begin{split}
|(H_\lambda^\omega - z)^{-1}(n,m)|^s \leq  
|(H_\lambda^\omega(S) - z)^{-1}(n,m)|^s \\
+    \sum_{k \in S^c, l \in S} |(H_\lambda^\omega(S) - z)^{-1}(n,k)|^s
\\
|[(H_0 -
H_{0,S^c})(k, l)]|^s 
|(H_\lambda^\omega(S) - z)^{-1}(l,m)|^s  
\end{split}
\end{equation}
is valid.  Observe that since $k \in S^c$ and $l\in S$, we have
$$
(H_\lambda^\omega(S) - z)^{-1}(k,l) = (H_{0,S^c} -z)^{-1}(k,l),
$$
since by assumption (\ref{ass1}) $|E| > \|H_0\|_s$ hence
$|E| > \|H_{0,S^c}\|_s$ .  Therefore
the estimates of proposition (\ref{aprop4}) are valid and after taking
averages in the above inequality we have
\begin{equation}
\begin{split}
\E \left\{|(H_\lambda^\omega - E -i\epsilon)^{-1}(n,m)|^s\right\} \leq  
|(H_{0,S^c} - E -i\epsilon)^{-1}(n,m)|^s \\
+    \sum_{k \in S^c, l \in S} |(H_{0,S^c} - E -i\epsilon)^{-1}(n,k)|^s
\\
|[(H_0 -
H_{0,S^c})(k, l)]|^s 
\E \left\{ |(H_\lambda^\omega(S) - E -i\epsilon)^{-1}(l,m)|^s \right\}\\
\leq \frac{1}{dist([-\|H_0\|_s, \|H_0\|_s,E)} \\
+  \|H_0\|_s^s C(H_{0,S^c}, E, s) 
sup_{l \in S} \E \left\{ |(H_\lambda^\omega(S) - E -i\epsilon)^{-1}(l,m)|^s 
\right\}\\
\end{split}
\end{equation}
where the constant $C(H_{0,S^C}, E , s)$ is given by proposition
(\ref{aprop4}), and is finite for each $E \in \RR \setminus
[-\|H_0\|_s, \|H_0\|_s]$, fixed.
Since now both $l, m$ are in S, we can use the estimate in Case 1 to
conclude that,
\begin{equation}
\begin{split}
\E \left\{|(H_\lambda^\omega - E -i\epsilon)^{-1}(n,m)|^s\right\} \\
\leq  
1/dist([-\|H_0\|_s, \|H_0\|_s],E) +  \|H_0\|_s^s C(H_{0,S^c}, E, s)
 D_0(E, \lambda, s) 
\end{split}
\end{equation}
We designate the quantity on the right hand side of the above inequality
as $D_1(E, \lambda, s)$.

{\noindent \it Case 3}: In case 2, we started with a uniform bound
valid for the average of the s-th moment when the sites were both on S,
to get a E dependent bound, but uniform in m, n when at least one of
them is in S.  We thus could relax the condition on m,n at the cost of
having the bounding constant depend on E.  It is clear that we can
repeat this trick, to cover all sites m, n in $\ZZ^\nu$.
Therefore using the result proved in Case 2 above we
repeat the proof of case 2 when $n, m \in S^c$ and we set the
resulting constant in the inequality as $D_2(E,\lambda, s)$.  Now we
take 
$$
D(E, \lambda, s) = max \left\{D_0(E, \lambda, s), D_1(E, \lambda, s), 
D_2(E,\lambda,s)\right\}.
$$
With this constant the proposition is valid.  Further we see from the
proof that each
of the $D_i(E, \lambda, s)$ is uniformly bounded in any compact subset
of $\rho(H_0)$ hence $D(E, \lambda, s)$ also satisfies this property. 

{\noindent \bf Remark:} The estimate in the above proposition did not
depend upon the set S, so if we set the potential to be zero at some,
or even all,  of
the sites in S, the result is still valid with the same bound.  Of course,
this amounts to shrinking the set S and 
when $S = \emptyset$, then the trivial uniform bound in terms of the
inverse of the distance of E to $[-\|H_0\|_s, \|H_0\|_s]$ is valid.

{\noindent \bf Proof of theorem (\ref{thm2}):}
The proof of item (1) of this theorem proceeds similar to that of item
(1) in the earlier theorem.  For that proof to go through we need
that 
$$
\sigma ~~ \int dt   \|A P_S e^{-itH_0}\phi\|   <\infty
$$
which we ensured by assumption of sparseness of S relative to $H_0$
with weight A.

(2) We prove this part on the absence of continuous spectrum outside 
$[-\|H_0\|_1, \|H_0\|_1]$, by using the estimates of
Aizenman-Molchanov.  We do the proof in two steps.  Step one consists
of noting that if the average Green function is bounded for each E,
then the sum of any finite number of them is also bounded.  Therefore
we need to look at the decay of the average Green function outside a
finite set of sites.  We determine the finite set based on the number
$0 < s < 1$ and proceed to show exponential decay on the complement of
that  set. 

Let $\Lambda_s(n)$ be the smallest cube centered at
n such that 
$$
B = \inf_{m \in \Lambda_s(n)^c \cap S}|a_m|^s(1 - s)^s D(s) /\|H_0\|_s^s > 1.
$$ 
where $D(s)$ is the constant appearing in the proposition
(\ref{aprop11}). Since $|a_m| \rightarrow \infty$ as $|m| \rightarrow \infty$,
such a cube exists for each fixed s in (0, 1) and each fixed n $\in
\ZZ^\nu.$  Then for each $m \in \Lambda_s(n)^s\cap S$ we have the estimate, as
in the proof of theorem (\ref{thm1})(2), 
\begin{equation}
\begin{split}
B \|H_0\|_s^s \E \{ |G(E+i\epsilon, n, m)|^s\} \leq \\ 
|a_m|^s(1 - s)^s D(s) \E \{ |G(E+i\epsilon, n, m)|^s\} \leq \\ 
\E \{ |(a_m V^\omega(m)\chi_S(m) - E -i\epsilon)
G(E+i\epsilon, n, m)|^s\} \leq \\
\delta_{n,m} + 
\sum_{k\in\ZZ^\nu} |\langle\delta_k, H_0 \delta_m\rangle|^s
\E \{|G(E+i\epsilon, n, k)|^s\}.
\end{split}
\end{equation}
and for E  in any compact subset of $ \RR \setminus [-\|H_0\|_s, \|H_0\|_s]$,
\begin{equation}
\begin{split}
|E|^s \E \{ |G(E+i\epsilon, n, m)|^s\} \leq \\ 
\E \{ |( - E -i\epsilon)
G(E+i\epsilon, n, m)|^s\} \leq \\
\delta_{n,m} + 
\sum_{k\in\ZZ^\nu} |\langle\delta_k, H_0 \delta_m\rangle|^s
\E \{|G(E+i\epsilon, n, k)|^s\}.
\end{split}
\end{equation}
for all $m \in \Lambda_s(n)^c \setminus S$.  Combining these two
estimates  we have that for any $m \in \Lambda_s(n)^c$, 
\begin{equation}
\begin{split}
 C(E,s) \E \{ |G(E+i\epsilon, n, m)|^s\} \leq \\ 
  \E \{ |(a_mV_S^\omega(m) - E -i\epsilon)
G(E+i\epsilon, n, m)|^s\} \leq \\
 \delta_{n,m} + 
\sum_{k\in \ZZ^\nu}
 |\langle\delta_k, H_0 \delta_m\rangle|^s
\E \{|G(E+i\epsilon, n, k)|^s\}.
\end{split}
\end{equation}
where we have made use of the fact that outside S, $V_S^\omega = 0$
and have set
$$
C(E, s) = 
\begin{cases} |E|^s, & ~~ m \notin S ~~ \text{and}\\
  B\|H_0\|_s^s & 
~~ m \in S \cap \Lambda_s^c.
\end{cases}
$$
Then by assumption on E and definition of B we have that $C(E,s) >
\|H_0\|_s^s$.  Using this fact and the proposition (\ref{aprop3}) 
we have the inequality, for each $\epsilon > 0$,
\begin{equation}
\begin{split}
 C(E, s) \sum_{m \in \Lambda_s(n)^c} \E \{ |G(E+i\epsilon, n, m)|^s\} \leq \\ 
\sum_{m \in \Lambda_s(n)^c} \delta_{n,m} + 
\sum_{m \in \Lambda_s(n)^c} \sum_{k \in \Lambda_s(n)^c}
 |\langle\delta_k, H_0 \delta_m\rangle|^s \E \{|G(E+i\epsilon, n, k)|^s\}
\\ 
+ \sum_{m \in \Lambda_s(n)^c} \sum_{k \in \Lambda_s(n)} 
 |\langle\delta_k, H_0 \delta_m\rangle|^s \E \{|G(E+i\epsilon, n, k)|^s\}.
\end{split}
\end{equation}
Using this estimate, we get
the bound,
\begin{equation}
\begin{split}
\sum_{m \in \Lambda_s^{n}} \E \{ G(E+i\epsilon, n, m)|^s\} \\ \leq
 \sum_{j = 0}^\infty k_s^j  + (\sum_{l = 1}^\infty l^{\nu -1}k_s^l)
 ( \sup_{m} D(E, \sqrt{a_n a_m}, s))(1 + L_s(n)) < \infty,  
\end{split}
\end{equation}
where $K_s, k_s$ are defined as in the equation (\ref{defks}) with the
constant $C(E,s)$ given above replacing $C(E, \lambda, s)$ there and
the number $L_s(n)$ is the cardinality of the set $\Lambda_s(n)$,
which is finite by the assumption on $\{a_n\}$ and so is the number
$\sup_m D(E, \sqrt{a_n a_m}, s)$.  The above sum
converges by the assumption that $C(E,s) > \|H_0\|_s^s$, so that $k_s <
1$.  

From this estimate we conclude, as in the earlier theorem, that
$$
\PP\{\omega : \sigma_c(H^\omega) \subset [-\|H_0\|_s, \|H_0\|_s\} =
1.
$$
So taking a sequence $s_k \uparrow 1$, we see that with probability 1,
$$
\sigma_c(H^\omega) \subset \cap_k [-\|H_0\|_{s_k}, \|H_0\|_{s_k}]
 = [-\|H_0\|_{1}, \|H_0\|_{1}]
$$
since a countable intersection of sets of measure 1 also has 
measure 1.

\section{Examples}

In this section we present a general class of examples of operators $H_0$ and
subsets S of $\ZZ^\nu$ that satisfy our assumptions.
The examples for $H_0$ comes from  the spectral
representation of $\Delta$.

Let $H_0$ be the operator of multiplication by a function  h in
the spectral representation of $\Delta$, where
\begin{assum}
\label{ass3}
\begin{enumerate}
\item h is a real valued 
$C^{2\nu + 2}$ function on $[0, 2\pi]^{\nu}$ with 
\begin{equation}
C_h \equiv \sup_{\alpha} \sup_{\theta}
 |\frac{\partial^{\alpha}}{\partial\theta^\alpha}h(\theta)|  <
\infty
\end{equation}
where $\alpha$ is a multi index $(\alpha_1,\cdots, \alpha_\nu)$
with $\alpha_i \geq 0, ~~ \sum\alpha_i = 2\nu + 2$.
Assume further that 
$$
h^{(\alpha)}(\theta_1,\cdots,\theta_{i-1},0,\theta_{i+1},\cdots,
\theta_\nu)=
h^{(\alpha)}(\theta_1,\cdots,\theta_{i-1},2\pi,\theta_{i+1},\cdots,\theta_\nu)
$$
for each i=1,..,$\nu$ and each multi index $\alpha$ with
$|\alpha_i| \leq 2\nu +2$.

\item  h is separable, i.e. $h(\theta_1,\cdots, \theta_\nu) =
\sum_{i = 1}^\nu h_i(\theta_i)$.

\item For each $i=1,\cdots,\nu$,
$\frac{d^3}{d\theta_i^3}h_i(\theta_i) \neq 0$ whenever
$\theta_i$ is a zero of $\frac{d^2}{d\theta_i^2}h_i(\theta_i) =
0$ whose number is assumed to be finite.
\end{enumerate}
\end{assum}

{\noindent \bf Remark:} It may appear strange to the reader that we
need the separability condition (2) above.  The reason is that the
higher dimensional version of the proposition (\ref{aprop3})
( on stationary phase)  is not known when the second
derivative of the phase function is singular at some point. 

\begin{prop}
\label{prop2}
Let h be a function as in assumption (\ref{ass3})(1).
Fix any $s_0 > \nu/2\nu + 1$.
Then there is a constant $C(s_0, C_h)$ such that 
for any $s_0 < s$, 
$$
\sup_{n \in \ZZ^\nu} \sum_{m \in \ZZ^\nu} |\langle \delta_n,
H_0\delta_m\rangle|^{s} < C(s_0, C_h). 
$$
\end{prop}

{\bf Proof:}
Writing the expression for $\langle \delta_n,
H_0\delta_m\rangle$ in the spectral representation for $H_0$
we have
\begin{equation}
\langle \delta_n, H_0\delta_m\rangle = \frac{1}{(2\pi)^\nu}\int_{[0,2\pi]^\nu}
e^{-i\sum_{j=1}^\nu(n-m)_j\theta_j}h(\theta_1,\cdots, \theta_\nu)
\prod_{j=1}^\nu d\theta_j
\end{equation}
Now
using assumption (\ref{ass3})(1), integration by parts (2$\nu$ +
1) times with respect to the co-ordinate $\theta_i$ which is chosen such
that $(n - m)_i \geq |n-m|/\nu$, gives the crude estimate
\begin{equation}
|\langle \delta_n, H_0 \delta_m\rangle| \leq C_h \frac{\nu^{2\nu +
1}}{|n-m|^{2\nu+1}}, ~~ n \neq m.
\end{equation}
This implies the proposition.  Here
the assumption on the derivatives at the boundary are made so that the
boundary terms in the integration by parts vanish at each stage.
In the following proposition we
denote by $\| h^\prime\| = \sup_{i} \sup_{\theta \in [0,2\pi]}
|h^\prime_i(\theta)|$.

\begin{prop}
\label{prop3}
Let h be a function satisfying assumption (\ref{ass3}).  Then we
have the following estimates.
\begin{enumerate}
\item $|\langle \delta_n, e^{\{-itH_0\}}\delta_m \rangle| \leq
C/|n-m|^{2\nu+1}, ~~ if ~~ \nu |t|\|h^\prime\|/|n-m| \leq 1/2,$ 
\item $|\langle \delta_n, e^{\{-itH_0\}}\delta_m \rangle| \leq
C/|t|^{\nu/3}, |t| \geq t_0$, for some $t_0$ large
where C, $t_0$ are independent of n,m.
\end{enumerate}
\end{prop}

{\bf Proof:}
The proof of the first estimate is a repeated integration by parts 
(see Stein, \cite{st}, VIII.1.3. Proposition 1)
applied to one of the integrals in the product
\begin{equation}
\begin{split}
\langle \delta_n, e^{\{-itH_0\}}\delta_m \rangle 
 = \prod_{i = 1}^\nu \frac{1}{2\pi}\int d\theta e^{-ith_i(\theta) + i
(n-m)_i\theta}. 
\end{split}
\label{prod}
\end{equation}
We use that integral (in the product) for which $|(n-m)_i|
\geq |n-m|/\nu$ to do the integration by parts.  
Our assumption on the equality of the derivatives at the
boundaries ensures that the boundary terms vanish for up to $2\nu+1$
derivatives, while the condition on $t$ and $n-m$ ensures that
$| 1 - t h^\prime_i(\theta)/(n-m)_i| > 1/2$, for all $\theta \in [0,
2\pi]$.  From this lower bound the estimate
$|[th^\prime_i(\theta)] - (n-m)_i| \geq |n - m|/2\nu$ is clear and that
this implies the estimate is straight forward.

To get the second estimate, 
we consider  one of the integrals , say the one corresponding to the
index i, in the product in equation
(\ref{prod}) and obtain a $C/t^{(1/3)}$ bound for all $t > t_0$, the
estimates for the other integrals is similar, resulting in the stated
bound of the Proposition.

We know by
assumption (\ref{ass3})(3), that the number of points in [0, 2$\pi$] where
the second derivative of $h_i$ vanishes is finite, say $x_1,\cdots, x_N$ 
and at these points the third derivative does not vanish.
We also know from assumption (\ref{ass3}), that $h_i$ is $C^{2\nu+2}$.
So we take any x in [0, 2$\pi$]
and expand $h_i$ about x 
using the Taylors formula with reminder to get 
$$
h_i(y) = \sum_{j=0}^{M} \frac{h_i^{(j)}(x)}{j!} (x-y)^j + R_i^{(M)}(y), ~~
0 \leq M \leq \nu +1,
$$
for y in a neghbourhood of x. 
We then consider the sets 
$$
S_1(x) = \left\{ y \in [0,2\pi] : |R_i^{(2)}(y)| < |h_i^{(2)}(x)|/2\right\}.
$$
when $x \notin \left\{x_1, \cdots, x_N\right\}$ and 
$$
S_1(x_j) = \left\{ y \in [0,2\pi] : |R_i^{(3)}(y)| < |h_i^{(3)}(x_j)|/2\right\},
j = 1, \cdots, N.
$$
Each of the sets $S_1(x)$ is relatively open in $[0, 2\pi]$,
by the continuity of the reminder terms $R_i^{(k)}$, k = 2,3. 
So one can find neighbourhoods $S(x)$ of x so that $\overline{S(x)}
\subset S_1(x)$.
Clearly $\cup_{x \in [0, 2\pi]} S(x)$ covers the (compact) set $[0,
2\pi]$.   Therefore, a finite collection of 
the above sets $S(x)$ cover $[0,  2\pi]$.  Let 
$S(\alpha_j), j = 1, \cdots, M$ cover $[0, 2\pi]$.
It is possible that some of the points $\alpha_j$ will correspond to
some $x_k$ at which the second derivative of $h_i$ vanishes. 
Let us index the $\alpha_j$ such that $\alpha_j \in \left\{x_k, k
=1,\cdots, N\right\}, j = 1, \cdots, K$ ( $K \leq N$ ) and the
remaining $\alpha_j$'s are points where the second derivative of $h_i$
does not vanish.
Let $\psi_j$ be a partition of unity subordinate to the cover
$S(\alpha_j), j= 1, \cdots, M$.

Then, 
\begin{equation}
 \int_0^{2\pi} d\theta e^{-ith_i(\theta) + i (n-m)_i\theta} = 
\sum_{j = 1}^M \int_0^{2\pi} d\theta e^{-ith_i(\theta) + i (n-m)_i\theta}
 \psi_j(\theta) 
\end{equation}
Suppose for the index j, the support of $\psi_j$ is contained in $(0,
2\pi)$.  Then 
 the estimate for the integral  
$\int_0^{2\pi} d\theta e^{-ith_i(\theta) + i (n-m)_i\theta}
\psi_j(\theta)$
follows from the proposition (\ref{aprop3}) where we set $\lambda
= t$ and $\phi(\theta) = h_i(\theta) + ((n-m)_i/t)\theta$. We
note that since the second and third derivatives of the $\phi$
above are independent of $t$, the proposition is still
applicable, even though it seems that \'a priori $\phi$ has a
``$\lambda$'' dependence. Then we
get $C/|t|^{1/3}$ bound for $j = 1, \cdots, K$ and $C/|t|^{1/2}$
bound for the remaining $j$s, for large enough $|t|$.

It is in general possible for an arbitrary $h_i$, integration by parts
leaves non-zero boundary terms at the points 0 and $2\pi$, 
so we deal with this case separately.

Suppose, $\psi_{j_1}$ and $\psi_{j_2}$ are the functions  
which have the points 0 and $2\pi$, respectively, in their support.  Then, we
first observe that we could have chosen the sets $S(\alpha_{j_1}) =
[0, \beta)$ and $S(\alpha_{j_2}) = (2\pi - \beta, 2\pi]$ for some
$\beta >0$ (to be the only ones containing 0 and $2\pi$ ).
The $\beta$ could be determined based on whether $h_i^{(2)}(0)$  is zero or
not and the associated reminder term in the Taylors formula with
reminder, by first extending $h_i$ to a periodic function in 
$C^{2\nu+2}(\RR)$ since by assumption $h_i^{(k)}(0) = h_i^{(k)}(2\pi),
k = 1, \cdots, 2\nu + 2$, at 0 (or equivalently at $2\pi$).
 Then we could have chosen
\begin{equation}
\psi_{j_1}(x) =
\begin{cases}
  & g(x) e^{-(x -  \beta)^{-2}}, ~~ x \in [0, \beta)\\
 &  0, ~~ \text{otherwise}
\end{cases}
\end{equation}   
and
\begin{equation}
\psi_{j_2}(x) =
\begin{cases}
& g(x) e^{-(x - 2\pi + \beta)^{-2}}, ~~ x \in (2\pi - \beta,
2\pi]\\
& 0, ~~ \text{otherwise}
\end{cases}
\end{equation}   
where $g(x)$ is the usual normalizing function to
get the partition of unity.  

We then consider the integral
$$
\int_0^{2\pi} d\theta e^{-ith_i(\theta) + i (n-m)_i\theta}
\{ \psi_{j_1}(\theta) + \psi_{j_2}(\theta)\} d\theta
$$
and it can be written as 
$$
\int d\theta e^{-ith_i(\theta) + i (n-m)_i\theta}
\{ \psi_{j_1}(\theta + 2\pi) + \psi_{j_2}(\theta)\} d\theta.
$$
Then the function $\phi(x) = \psi_{j_1}(x) + \psi_{j_2}(x)$ is smooth
and has support satisfying the assumptions of the proposition
(\ref{aprop3}) so that we can get a bound of $C/|t|^{1/3}$ or
$C/|t|^{1/2}$ for the above integral, based on the vanishing or
otherwise of the second derivative of $h_i$ at 0 (equivalently at $2\pi$).

\begin{lemma}
\label{lem4}
Let $\nu \geq 5$.  Let S be a subset of $\ZZ^\nu$ satisfying 
$|S \cap \Lambda| \leq |\Lambda|^\alpha, ~~ 0 < \alpha <
2(1/3 - 1/\nu)$, for any cube $\Lambda$ as $|\Lambda| \rightarrow
\infty.$  Let   
$H_0$ be the operator associated with the function $h$ satisfying the
assumptions (\ref{ass3}). Then
\begin{enumerate}
\item
S is sparse relative to $H_0$. 
\item 
If $\DD$ denotes the set of vectors of finite support in
$\ell^2(\ZZ^\nu)$, then for each t fixed
$\|(1 + |m|)^\beta e^{-itH_0}\phi\| < \infty$, for $\nu + 1 > \beta > \nu$.
\end{enumerate}
\end{lemma}
 
{\bf Proof:}
To show that S is sparse relative to $H_0$, we consider
$$
\|P_S e^{-itH_0}\phi\|
$$
for $\phi$ such that 
 $\langle \phi, \delta_k\rangle = 0$ for all but finitely many k
and $\|\phi\| =1$.
Since $H_0$ has purely absolutely continuous spectrum, under
assumption (\ref{ass3}) on h, this collection of $\phi$ forms a
dense subset of the absolutely continuous spectral space of
$H_0$.  We show that this quantity is integrable in t
$\geq 1$, for all m and the integral is bounded by a constant
independent of m and $\phi$.  We have 
\begin{equation}
\begin{split}
\int dt \|P_S e^{-itH_0}\phi\|  \leq \\  
\|\phi\|\int dt \left(\sum_{m \in S} \left(\sum_{n:\phi(n)\neq 0 }
|\langle \delta_m, e^{-itH_0}\delta_n\rangle|^2\right)\right)^{1/2} \\
\leq \|\phi\|\int dt \left( \left(\sum_{n : \phi(n) \neq 0}  \sum_{m \in S}
|\langle \delta_m, e^{-itH_0}\delta_n\rangle|^2\right)\right)^{1/2} \\
\leq \|\phi\|\int dt \left( \sum_{n : \phi(n) \neq 0} \left(\sum_{m \in S : |n -m| >
 2 \nu t\|h'\|}
|\langle \delta_m, e^{-itH_0}\delta_n\rangle|^2 \right. \right.
\\   
+ \left. \left. \sum_{m \in S : |n -m| \leq 2\nu t\|h'\|}
|\langle \delta_m, e^{-itH_0}\delta_n\rangle|^2 \right) \right)^{1/2} 
\end{split}
\end{equation}
The last two summands are estimated using the two estimates of
proposition (\ref{prop3}), to get 
\begin{equation}
\begin{split}
\int dt \|P_S e^{-itH_0}\phi\|  \leq \\  
\|\phi\|\int dt \left(\sum_{n : \phi(n) \neq 0}   
\left( \sum_{m \in S : |n -m| > 2\nu t\|h'\|} C/|n-m|^{2\nu} \right.
\right. \\ \left. \left. +
\sum_{m \in S : |n -m| \leq 2\nu t\|h'\|} C/|t|^{2\nu/3}\right)
\right)^{1/2}  \\
\leq \|\phi\|\int dt  \left(\sum_{n : \phi(n) \neq 0}  \left( C/|t|^{\nu-} 
 +
 C|t|^{\alpha\nu}/|t|^{2\nu/3}\right)
\right)^{1/2}  \\
< C \|\phi\| \#\{n: \phi(n) \neq 0\} < \infty, 
\end{split}
\end{equation}
in view of the assumptions on $\nu$ and $\alpha$ and the finiteness of
the support of $\phi$, with \# denoting the cardinality of the set.

Part (2) is a direct consequence of the finiteness of the
support of $\phi$ and the estimate in proposition (\ref{prop3})(1).

We take any subset S of $\ZZ^\nu$, satisfying the assumption in
lemma (\ref{lem4}).  Consider for any k $\in \ZZ^+$, the function $h(\theta) =
\sum_{i = 1}^\nu 2 \cos k\theta_i$, so that $H_0 = \sum_{i =
1}^\nu (T_i^k + T_i^{-k})$, $T_i$ denoting the shift by 1 in the
i -th direction in $\ZZ^\nu$.  $\Delta$ corresponds to $k =1$.
In this case S is $H_0$ sparse and in theorem (\ref{thm2})   
the mobility edges are $\{-2\nu, 2\nu\}.$

We presented in this paper a class of random operators, having
both absolutely continuous spectrum and dense pure point spectrum.
The a.c. spectrum seems to come from the fact
that mostly the potential is zero, while the dense pure point spectrum
seems to come from localization near the potential sites.
The interesting aspect of the result is that there need not be any
structure for S.  One only requires that the set be asymptotically 
sparse.  Our examples include cases where S is a subgroup of $\ZZ^\nu$, for
large $\nu$ and then  the results in this paper
also have examples of ergodic potentials (with
respect to S action) exhibiting the a.c spectrum and
dense pure point spectrum. 

The mobility edges are also
identified for a class of potentials and a class of free operators
provided the coupling constants go to infinity at infinity.  In the
paper of Kirsch-Krishna-Obermeit \cite{kko} we showed similar result
for the case when the coupling constants decay to zero.  Such examples
in addition to sparseness also can be included here and the proof 
goes through for that case also.

\section{Appendix:}

In the appendix we collect a few results for the convenience of
the reader.

In the paper \cite{am} Aizenman-Molchanov introduced the decoupling
principle, which was at the heart of their method of proving
localization.  The lower bounds that they obtain on the expected
values of some random variables together with a uniform bound on the 
low moments on the Green functions of the problem, made the proof
possible.  

Their decoupling, stated in a version relevant for this paper is the following.
The proof of this lemma is almost identical to the one when $S =
\ZZ^\nu$, whose proof can be found in either Aizenman-Molchanov \cite{am}
Aizenman-Graf \cite{ag}.  Nevertheless we present it for the
convenience of the reader.

\begin{prop}[Aizenman-Molchanov]
\label{aprop1}
Consider the operator $H^\omega_\lambda$ with $V_S^\omega(m)$
satisfying the assumptions(\ref{ass2}).  Then for any $\lambda
>0$, $0 < s <1$, there is a positive constant $D(s)$ depending only
upon $\mu$ and s, but bounded above and below
as $s \rightarrow 1$, such that for each  $m \in S$,
\begin{equation}
\begin{split}
|\lambda|^s (1 - s)^s D(s) \E\{|G(E+i\epsilon, n,m)|^s\} \\
\leq \E\{
|(\lambda V^\omega(m) - 
E - i\epsilon)G(+i\epsilon,n,m)|^s\},
\end{split}
\end{equation} 
for any $n \in \ZZ^\nu$.
\end{prop}

{\noindent \bf Proof:}  First we note that our assumption on the
measure $\mu$ (the distribution of the random variables
$V_S^\omega(n)$ for any $n \in S$), is 1 regular in the sense of
Aizenman-Molchanov \cite{am}.  Therefore the
lemmas III.1 and III.2 of Aizenman-Molchanov \cite{am}, tell us
that the measures 
$ d\mu_s(x) = |x - \alpha|^s  d\mu(x)$ are respectively are 1 and
$ d\mu_s(x) = |x - \beta|^{-s} d\mu(x)$ are respectively are 1 and
(1-s) regular.  Their
proofs of the lemmas III.1 and III.2  applied to prove their lemma
3.1(i) show that we have the estimate,
$$
\int |x - \eta|^s |x - \beta|^s d\mu(x) \geq (k_s)^s \int |x -
\beta|^s d\mu(x)
$$  
with the constant $k_s = (1 - s)/ D(s)^s$, D(s) a constant
depending only upon the measure $\mu$, s but independent of 
$\eta, \beta \in \CC$ and also bounded above and below as $s \rightarrow 1$. 
Using this estimate we see that for any real number
$\gamma$,
$$
\int |\gamma x - \eta|^s |\gamma x - \beta|^s d\mu(x) \geq 
|\gamma|^s(k_s)^s \int |\gamma x - \beta|^s d\mu(x)
$$  
  
Once this estimate is in place, the proof of the proposition is as in
the proof of the decoupling lemma (2.3) of Aizenman-Molchanov
\cite{am}.  Finally we remark that the resond we needed $m \in S$ is
that otherwise $V_S^\omega(m) =0$ and there is no random variable to
integrate!  This was essential in getting the uniform bounds on the
energy E.

As an immediate corollary we see that 

\begin{prop}
\label{aprop11}
Consider the operator $H^\omega_\lambda$ with $V_S^\omega(m)$
satisfying the assumptions of theorem (\ref{thm2}).  Then 
there is a positive constant $D(s)$ depending only
upon $\mu$ and s, but bounded above and below
as $s \rightarrow 1$, such that for each  $m \in S$,
\begin{equation}
\begin{split}
|a_m|^s (1 - s)^s D(s) \E\{|G(E+i\epsilon, n,m)|^s\} \\
\leq \E\{
|(\lambda V^\omega(m) - 
E - i\epsilon)G(+i\epsilon,n,m)|^s\},
\end{split}
\end{equation} 
for any $n \in \ZZ^\nu$.
\end{prop}

{\noindent \bf Proof:}
This is an easy application of the proof of the earlier proposition
where for each $m \in S$ instead of $\lambda$ we now have $a_m$ as the
coupling constant.

\begin{prop}
\label{aprop4}
Let  $B$ be a bounded self-adjoint operator, commuting with $\Delta$
and satisfying the assumption (\ref{ass1}) for some $1 > s_0 >0$.
For each $s \in (s_0, 1)$, let $E \in [-\|H_0\|_s, \|H_0\|_s]$  
$$
\sum_{m \in \ZZ^\nu} |(B - z)^{-1}(n,m)|^s \leq C(B,E,s) < \infty. 
$$ 
with $Re(z) = E$.  The constant $C(B, E, s)$ is bounded as a function
of E on any compact subset of $\RR \setminus [-\|H_0\|_s, \|H_0\|_s]$.
\end{prop}

{\noindent \bf Proof:} Under the assumptions on z, it is in the
resolvent set of B and the bounded operator
$(B - z)^{-1}$ can be expanded using the Neumann series, which
converges under the assumption on E = Re(z).  Therefore we consider 
$$
(B - z)^{-1}(n,m) = \frac{1}{z}\sum_{k = 0}^\infty \langle \delta_n ,
\frac{H_0^k}{z^k} \delta_m \rangle.
$$
Taking the absolute values to the power s and using the inequality 
$\sum |x_i|^s \geq (|\sum x_i|)^s$, for the given s, we get that
$$
|(B - z)^{-1}(n,m)|^s \leq \frac{1}{|z|}\sum_{k = 0}^\infty |\langle \delta_n ,
\frac{H_0^k}{z^k} \delta_m \rangle|^s.
$$
Therefore for any cube $\Lambda$ centered at $0$ in $\ZZ^\nu$, we have
$$
\sum_{m \in \Lambda}|(B - E)^{-1}(n,m)|^s \leq \sum_{m \in \Lambda}
\frac{1}{|z|}\sum_{k = 0}^\infty |\langle \delta_n ,
\frac{H_0^k}{z^k} \delta_m \rangle|^s,
$$
which implies
$$
\sum_{m \in \Lambda}|(B - E)^{-1}(n,m)|^s \leq
\frac{1}{|z|}\sum_{m \in \Lambda} \sum_{k = 0}^\infty |\langle \delta_n ,
\frac{H_0^k}{z^k} \delta_m \rangle|^s.
$$
On the other hand from the assumption on $B$ we have that for any
positive integer k, 
\begin{equation}
\begin{split}
\langle \delta_n ,
\frac{H_0^k}{z^k} \delta_m \rangle 
= \frac{1}{z^k} \sum_{l_1, \cdots, l_{k-1}}
 \langle \delta_n , B \delta_{l_1}\rangle 
 \langle \delta_{l_1} , B \delta_{l_2}\rangle, 
\cdots , \\ 
 \langle \delta_{l_{k-2}} , B \delta_{l_{k-1}}\rangle 
 \langle \delta_{l_{k-1}} , B \delta_{m}\rangle. 
\end{split}
\end{equation}
Therefore estimating after taking the sum over $\Lambda$ we get, since
$|z| \geq |E|$,
\begin{equation}
\begin{split}
\sum_{m \in \Lambda} |\langle \delta_n ,
\frac{H_0^k}{z^k} \delta_m \rangle|^s 
= \frac{1}{|E|^{sk}} \sum_{m \in \Lambda}\sum_{l_1, \cdots,
l_{k-1}}
 |\langle \delta_n , B \delta_{l_1}\rangle |^s
 |\langle \delta_{l_1} , B \delta_{l_2}\rangle|^s, 
\cdots , \\ 
 |\langle \delta_{l_{k-2}} , B \delta_{l_{k-1}}\rangle|^s 
 |\langle \delta_{l_{k-1}} , B \delta_{m}\rangle|^s. 
\end{split}
\end{equation}
This results in 
\begin{equation}
\begin{split}
\sum_{m \in \Lambda} |\langle \delta_n ,
\frac{H_0^k}{z^k} \delta_m \rangle|^s 
= \frac{1}{|E|^{sk}} \|B\|_s ^{s(k -1)}\sup_{l_{k-1} \in \ZZ^\nu}
 \sum_{m \in \Lambda}
 |\langle \delta_{l_{k-1}} , B \delta_{m}\rangle|^s\\
\leq \frac{\|B\|_s^{sk}}{|E|^{sk}}. 
\end{split}
\end{equation}
This implies that 
$$
\sum_{m \in \Lambda}|(B - z)^{-1}(n,m)|^s \leq 
\frac{1}{|E|}\sum_{k = 0}^\infty \frac{\|B\|_s^{sk}}{|E|^{sk}} < \infty , 
$$
under the assumptions on E, with the sum on the right denoted C(B, E, s).
Since the bound on the right is independent of $\Lambda$, it is also
valid for the supremum over all such $\Lambda$ and by taking a
collection of cubes increasing to $\ZZ^\nu$,  we conclude the
proposition.

We finally restate the proposition on stationary phase estimate
from Stein \cite{st}, VIII.1.3., proposition 3.  Below $\phi$ is
a real valued function having (k+1) continuous derivatives in (a,b).
and $\psi$ is a smooth function whose support contains only one
critical point of $\phi$.  We note that the
assumptions on $\phi$ below allow us to approximate it by
$(x-x_0)^k [ \phi^{(k)}(x_0) + \epsilon(x)]$ with
$\|\epsilon(x)\|_{\infty} \leq \phi^{(k)}(x_0)/2$  in the
support of $\psi$, if it is small enough,
using the Taylor`s theorem with reminder.  
These are the conditions on $\phi$ and $\psi$ required in the proof of
the proposition below.  
\begin{prop}[Stein]
\label{aprop3}
Suppose $k \geq 2$, and 
$$
\phi(x_0) = \phi^\prime(x_0) = \cdots = \phi^{(k-1)}(x_0) = 0, 
$$
while $\phi^{(k)}(x_0) \neq 0$.  If $\psi$ is supported in a
sufficiently small neighbourhood of $x_0$, then
$$
I(\lambda) = \int ~~ e^{i\lambda \phi(x)} \psi(x) dx \approx
{\lambda}^{-1/k} \sum_{j=0}^\infty a_j {\lambda}^{-j/k},
$$
in the sense that, for all integers N and r,
$$
\frac{d^r}{dx^r}\left[ I(\lambda) - {\lambda}^{-1/k}\sum_{j=0}^N
a_j {\lambda}^{-j/k}\right] = O({\lambda}^{-r-(N+1)/k}) ~~ as ~~
\lambda \rightarrow \infty.
$$ 
\end{prop}

\end{document}